\documentclass[aps,prl,preprint,superscriptaddress]{revtex4}

\usepackage{graphicx}
\usepackage{amsmath}
\usepackage{amsfonts}

\usepackage{graphicx}
\usepackage{dcolumn}
\usepackage{bm}
\usepackage{amssymb}

\begin{document}

\title{Control of Mooij correlations at the nanoscale \\ in the disordered metallic Ta\,--\,nanoisland FeNi multilayers}

\author{N.~N. Kovaleva} 
\affiliation{Department of Physics, Loughborough University,
LE11 3TU Loughborough, United Kingdom}
\affiliation{P.N. Lebedev Physical Institute, Russian Academy of Sciences, 119991 Moscow, Russia}
\author{F.~V. Kusmartsev}
\affiliation{Department of Physics, Loughborough University,
LE11 3TU Loughborough, United Kingdom}
\affiliation{Micro/Nano Fabrication Laboratory (MNFL), Microsystem and Terahertz Research Center, 610200 Chengdu, China}
\author{A.~B. Mekhiya}
\affiliation{P.N. Lebedev Physical Institute, Russian Academy of Sciences, 119991 Moscow, Russia}
\author{I.~N. Trunkin}
\affiliation{National Research Center, Kurchatov Institute, 123182 Moscow, Russia}
\author{\mbox{D. Chvostova}}
\affiliation{Institute of Physics, Academy of Sciences of the Czech Republic, 18221 Prague, Czech Republic}
\author{A.~B. Davydov}
\affiliation{P.N. Lebedev Physical Institute, Russian Academy of Sciences, 119991 Moscow, Russia}
\author{L.~N. Oveshnikov}
\affiliation{P.N. Lebedev Physical Institute, Russian Academy of Sciences, 119991 Moscow, Russia}
\affiliation{National Research Center, Kurchatov Institute, 123182 Moscow, Russia}
\author{O. Pacherova} 
\affiliation{Institute of Physics, Academy of Sciences of the Czech Republic, 18221 Prague, Czech Republic}
\author{I.~A. Sherstnev} 
\affiliation{P.N. Lebedev Physical Institute, Russian Academy of Sciences, 119991 Moscow, Russia}
\author{A. Kusmartseva}
\affiliation{Department of Physics, Loughborough University,
LE11 3TU Loughborough, United Kingdom}
\author{K.~I. Kugel}
\affiliation{Institute for Theoretical and Applied Electrodynamics, Russian Academy of Sciences, 125412 Moscow, Russia} 
\affiliation{National Research University Higher School of Economics, 101000 Moscow, Russia\\
E-mail: N.Kovaleva@lboro.ac.uk}
\author{A. Dejneka}
\affiliation{Institute of Physics, Academy of Sciences of the Czech Republic, 18221 Prague, Czech Republic}
\author{F. A. Pudonin}
\affiliation{P.N. Lebedev Physical Institute, Russian Academy of Sciences, 119991 Moscow, Russia}
\author{Y. Luo}
\affiliation{Micro/Nano Fabrication Laboratory (MNFL), Microsystem and Terahertz Research Center, 610200 Chengdu, China}
\author{B. A. Aronzon}
\affiliation{P.N. Lebedev Physical Institute, Russian Academy of Sciences, 119991 Moscow, Russia}

\date{\today}
 
\pacs{71.20.Lp 71.27.+a 78.20.-e}

\maketitle

{\bf Localisation phenomena in highly disordered metals close to 
the extreme conditions determined by the Mott-Ioffe-Regel (MIR) 
limit when the electron mean free path is approximately equal 
to the interatomic distance is a challenging problem. Here, to shed 
light on these localisation phenomena, we studied the dc transport 
and optical conductivity properties of nanoscaled multilayered films 
composed of disordered metallic Ta and magnetic FeNi nanoisland layers, 
where ferromagnetic FeNi nanoislands have giant magnetic moments of 
10$^3$--10$^5$ Bohr magnetons ($\mu_{\rm B}$). In these multilayered 
structures, FeNi nanoisland giant magnetic moments are interacting 
due to the indirect exchange forces acting via the Ta electron subsystem. We discovered that the localisation phenomena in the disordered Ta layer lead to a decrease in the Drude contribution of free charge carriers and the appearance of the low-energy electronic excitations in the 1--2 eV spectral range characteristic of electronic correlations, which may accompany the formation of electronic inhomogeneities.
From the consistent results of the dc transport and optical studies we 
found that with an increase in the FeNi layer thickness across the 
percolation threshold evolution from the superferromagnetic to 
ferromagnetic behaviour within the FeNi layer leads to the delocalisation of Ta electrons from the associated localised electronic states. 
On the contrary, we discovered that when the FeNi layer is discontinuous and represented by randomly distributed superparamagnetic FeNi nanoislands, the Ta layer normalized dc conductivity falls down below the MIR limit 
by about 60\%. The discovered effect leading to the dc conductivity fall below the MIR limit can be associated with non-ergodicity and purely quantum (many-body) localisation phenomena, which need to be challenged further.}

\vspace{1cm}
The phenomenon of negative temperature coefficient of resistivity (TCR) 
characterising linearly temperature-dependent resistivity variation 
$\rho(T)=\rho_0\left[ 1+\alpha_0(T-T_0) \right]$ is found in many 
highly disordered metals, amorphous metals, and metallic glasses in the 
form of both bulk solids and thin films. There the TCR ($\alpha_0$) values tend to be more negative with increasing resistivity $\rho_0$ as demonstrated by the numerous data collected in the original Mooij plot \cite{Mooij}. 
This phenomenon is beyond the ordinary Boltzmann theory of transport 
in metals, which rigorously requires that $d\rho/dT>0$. 
The original Mooij plot indicates that TCR changes sign in a relatively narrow range of resistivities $\rho_c$\,$\simeq$\,100--150\,
$\mu\Omega\cdot$cm. However, subsequently, it was argued by Tsuei 
that the original Mooij plot ($\alpha_0$ vs $\rho_0$) is not universal \cite{Tsuei}. It was shown that for the three-dimensional (3D) case $\rho_c$ depends on the material characteristics, Fermi surface wave vector $k_{\rm F}$ 
and the elastic mean free path $l_e$ of conduction electrons. 
More than 500 data points were collected from the literature 
by Tsuei in his seminal work \cite{Tsuei} indicating that $\rho_c$ 
spans from $\sim$\,30 to $\sim$\,300\,$\mu\Omega\cdot$cm in the 
Mooij plot coordinates.

Different microscopic scenarios have been developed to explain 
the origin of Mooij correlations in disordered metals. 
Following the chronological sequence, earlier the TCR and $\rho(T)$ 
were fairly well described using the physical concepts that have been 
put forward for liquid metals \cite{Ziman}. Some of them 
(for example, Hg and liquid Te) were proposed by Mott \cite{Mott1} 
as candidates for a metal at the borderline of the nonconducting 
behavior. Indeed, they demonstrate an intriguing behaviour where 
optical conductivity $\sigma_1(\omega)$ displays a minimum at zero 
frequency accompanied by a transfer of the oscillator strength to 
higher frequencies in the form of an infrared peak \cite{Smith}. 
Later on, the theory developed for liquid metals was extended for 
metallic glasses \cite{Nagel,JackleandFrobose}. A principal 
possibility of using the latter approach is based on the 
suggestion that the elements of the crystal structure are 
preserved in liquid and amorphous metals. The change in the 
atomic static structure factor characteristic of the liquid 
as a function of temperature was calculated including the 
usual Debye-Waller factor and the phonon contribution.

One popular opinion on the possible cause for the negative TCR 
phenomenon \cite{Kavech,Tsuei,Imry1,Gantmakher1,Gantmakher} invokes 
the ``weak localisation scenario'' and the formation of the impurity-induced bound electronic state, viz., the so-called Anderson localisation \cite{Anderson}. The related quantum-interference processes, or weak localisation corrections \cite{Gorkov,Abrikosov}, provide a consistent theory for the 
description of the low-temperature dc transport in metals at weak 
disorder. The ``weak localisation'' scenario does not assume any significant rearrangement and gap opening in the electronic spectra. In this case, 
the dc transport will be determined by the competition between the degradation of the quantum-interference effects as a result of inelastic scattering and the conventional Boltzmann electron transport in metals. 
Metals and alloys remain metallic with the resistivity lower than the critical $\rho^*$ value corresponding to the Mott-Ioffe-Regel (MIR) limit \cite{Ioffe,Mott,Gurvitch}
\begin{eqnarray}
\rho^*=\frac{\hbar k_{\rm F}}{ne^2l_e}=\frac{\hbar}{e^2}\frac{1}{k_{\rm F}}\sim 300\,\mu\Omega\cdot \rm{cm},
\label{MIR}
\end{eqnarray}
where $\hbar$ is the Planck's constant. 
The universality of the Mooji correlations found in 
highly disordered metals, which suggests an approximately 
linear correspondence between TCR and resistivity $\rho_0$ 
($\alpha_0$ vs $\rho_0$), where $\rho_0$ is referred to 
the room temperature conditions \cite{Mooij}, is striking. The Mooij correlations found in the high-temperature regime where 
incoherent thermal phonon excitations abound seems to be highly 
improbable. However, since under the condition of increased disorder 
in high-resistivity alloys scattering by static structural defects may 
occur more frequently than inelastic scattering by phonons, 
long-distance phase coherence required for interference processes 
may persist in the appropriate temperature range 
\cite{Kavech,Tsuei,Imry1,Gantmakher1,Gantmakher,Lee}.

The most recently developed model, which is unrelated to the 
Anderson localisation mechanism, was suggested by Ciuchi {\it et al.} \cite{Ciuchi1}. The origin of the Mooij correlations is associated with a 
strong enhancement of polaronic effects due to disorder \cite{Ciuchi1,Ciuchi2}. It is suggested that the reduced mobility of electrons in poor 
metallic conductors may allow the lattice deformations to self-trap 
electrons through a disorder-assisted polaronic effect 
\cite{Anderson1}. The mechanism may acquire a dominant role at sufficiently strong disorder leading to gap formation and substantial transfer 
of spectral weight (SW) away from the Fermi energy to the 
renormalised polaronic states.

The presented above theoretical 
models questioning the origin of Mooij correlations 
consider averaged characteristics which may be accurate 
in description of only homogeneous systems. However, inhomogeneities 
may be peculiar for disordered metallic systems. For example, 
the formation of additional defects under irradiation of an amorphous 
Pd$_{80}$Si$_{20}$ alloy with fast neutrons may lead to the formation 
of inhomogeneous structure, consisting of clusters 10--20 \AA\ in 
diameter with an enhanced electronic density surrounded by neighbouring 
regions with devastated electronic density \cite{Doi}. In addition, 
in the limit of strong localisation conditions when electronic wave 
functions become disturbed practically at each atomic site, strong 
electronic correlations should be taken into account. 
In general, the formation of electronic inhomogeneities should 
necessarily lead to the rearrangement of optical SW due to 
electronic correlations. 
The situation may resemble that discovered in the 
Kondo-lattice metal Tb$_2$PdSi$_3$, 
where the formation of electronic inhomogeneities \cite{Kugel} 
manifests itself in the optical conductivity with opening of 
the pseudogap and the appearance of the Mott-Hubbard-like electronic 
excitations capturing the shifted optical SW (see \cite{Kovaleva_Scirep} and the theoretical model presented in the Supplementary online 
information to the paper \cite{Kovaleva_Scirep}). 
Here electronic correlations inevitably existing between 
localised electrons \cite{Kovaleva_PRL,Kovaleva_PRB_LMO,Kovaleva_PRB_YTO,Oles} may develop when metallic charge carriers become self-localised in 
the metallic magnetic clusters near the large localised moments of 
Tb 4f states. This is also 
supported by the recent evidence of the formation of intrinsic 
inhomogeneous states represented by ferromagnetic clusters 
(called magnetic polarons) in magnetic materials, in which 
the electronic and magnetic properties are strongly modified by 
the exchange coupling between the conduction electrons and 
local magnetic moments \cite{Fisk}.

Recently, by using dc transport and wide-band spectroscopic ellipsometry experimental techniques we investigated disordered metallic $\beta$-Ta films \cite{Kovaleva_APL_1,Kovaleva_metals}, which are known to manifest negative TCR \cite{Read}. 
We discovered that with increasing degree of disorder the Drude 
contribution due to intraband absorption within the Ta 5d $t_{2g}$ 
band at the Fermi level decreases and simultaneously the higher-energy 
optical bands appear as satellites at around 2--4 eV, 
where the associated optical SW is recovered. 
The discovered optical SW transfer can hardly be associated 
with the mechanism proposed for liquid metals by Smith \cite{Smith}, 
because the spectral positions of the bands by far exceed 
the infrared frequencies, rather being characteristic for the systems 
with strong electron correlations \cite{Kovaleva_PRL,Kovaleva_PRB_LMO,Kovaleva_PRB_YTO}.

\begin{figure}[b]
\includegraphics*[width=10.5cm]{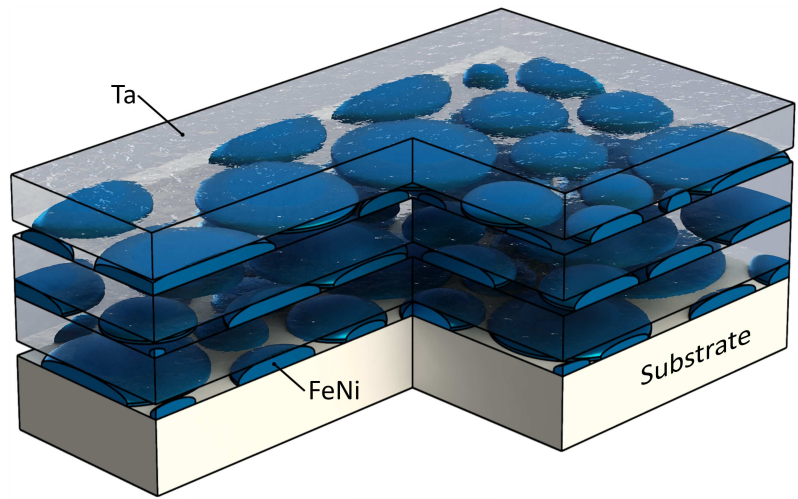}
\caption{{\bf A schematic picture of the multilayer film (MLF) samples 
(Ta--FeNi)$_{\rm N}$/Sitall substrate including nanoiland FeNi layers.}}
\label{MLF}
\end{figure}

To prove or disprove the existence of these phenomena, 
which can explain the origin of the Mooij correlations 
associated with the existence of nanoscale electronic 
inhomogeneities in highly disordered metals, 
one has to understand how to control them (disturb or disrupt). 
In the giant magnetoresistance (GMR) layered nanostructures 
the interlayer coupling between two ferromagnets is driven by the 
Ruderman-Kasuya-Kittel-Yoshida- (RKKY-) type \cite{Ruderman,Kasuya,Yoshida} indirect exchange interactions via itinerant charge carriers of a 
nonmagnetic spacer. Having this in mind, here we propose to apply 
the elaborated approach \cite{Kovaleva_APL_1,Kovaleva_metals} to 
study the dc transport and optical conductivity properties of the 
disordered metallic Ta interlayer in ultrathin multilayer structures 
Ta--Fe$_{21}$Ni$_{79}$ similar to those exhibiting the GMR effect. 
We investigated the rf-sputtered multilayer films (MLFs) 
(Ta--FeNi)$_{\rm N}$ in two limits of the FeNi layer thickness 
(i) when it is discontinuous and consists of inhomogeneously 
distributed single-domain ferromagnetic nanoislands having lateral 
sizes of 5--30\,nm \cite{Kovaleva_JNM} and possessing giant magnetic 
moments of 10$^3$--10$^5$ $\mu_{\rm B}$ (where $\mu_{\rm B}$ is 
the Bohr magneton) and (ii) when its nominal thickness is varied across 
the FeNi film percolation threshold around 1.5--1.8\,nm 
\cite{Sherstnev,Pudonin_JETP} (see the schematic picture of the MLF structure (Ta--FeNi)$_{\rm N}$ involving the nanoisland FeNi layer in Fig.\,\ref{MLF}). Here, with increasing the FeNi layer thickness across the percolation threshold the evolution from superparamagnetic (SPM) through superferromagnetic 
(SFM) to ferromagnetic (FM) behaviour within the FeNi layer 
\cite{Kleemann,Kovaleva_JNR,Lucinski} will modify the strength 
of the indirect exchange interaction between neighbouring FeNi 
layers. This, in turn, will strongly influence the 
itinerant charge carriers of the Ta layer promoting 
conditions for their enhanced localisation or delocalisation. 
It should be mentioned that, in addition to the intrinsic disorder 
of the Ta layer, its electron transport could strongly be affected 
by structural and magnetic inhomogeneities of nearby FeNi layers 
as it can give rise to a large-scale fluctuating potential of 
electrostatic and/or magnetic origin \cite{Aronzon1,Aronzon2}, 
leading to the appearance of topologically non-trivial spin structures \cite{Oveshnikov}.
Moreover, long-range many-body interactions between giant magnetic 
moments of inhomogeneously distributed FM FeNi nanoislands via
itinerant charge carriers of the disordered metallic Ta layer 
by means of RKKY-type indirect exchange could give rise to a substantial slowing down of all relaxation processes and to the effects of 
non-ergodicity and purely quantum (many-body) localisation effects 
\cite{Basko,Nandkishore,Huse,Khemani,Lukin}. Indeed, we discovered 
that when the FeNi layer is discontinuous and represented by 
randomly distributed distant FM FeNi nanoislands having giant 
magnetic moments, the Ta layer dc conductivity falls down below 
the MIR limit by about 60\% (normalised to the MIR limit value). 
The discovered phenomenon leading to the dc conductivity fall below 
the MIR limit need to be challenged further. The results of the present study could be important for probing the fundamental physics associated 
with the localisation phenomena in disordered metals and 
from the application point of view of the GMR effect.

\begin{figure}[b]
\includegraphics[width=12.5cm]{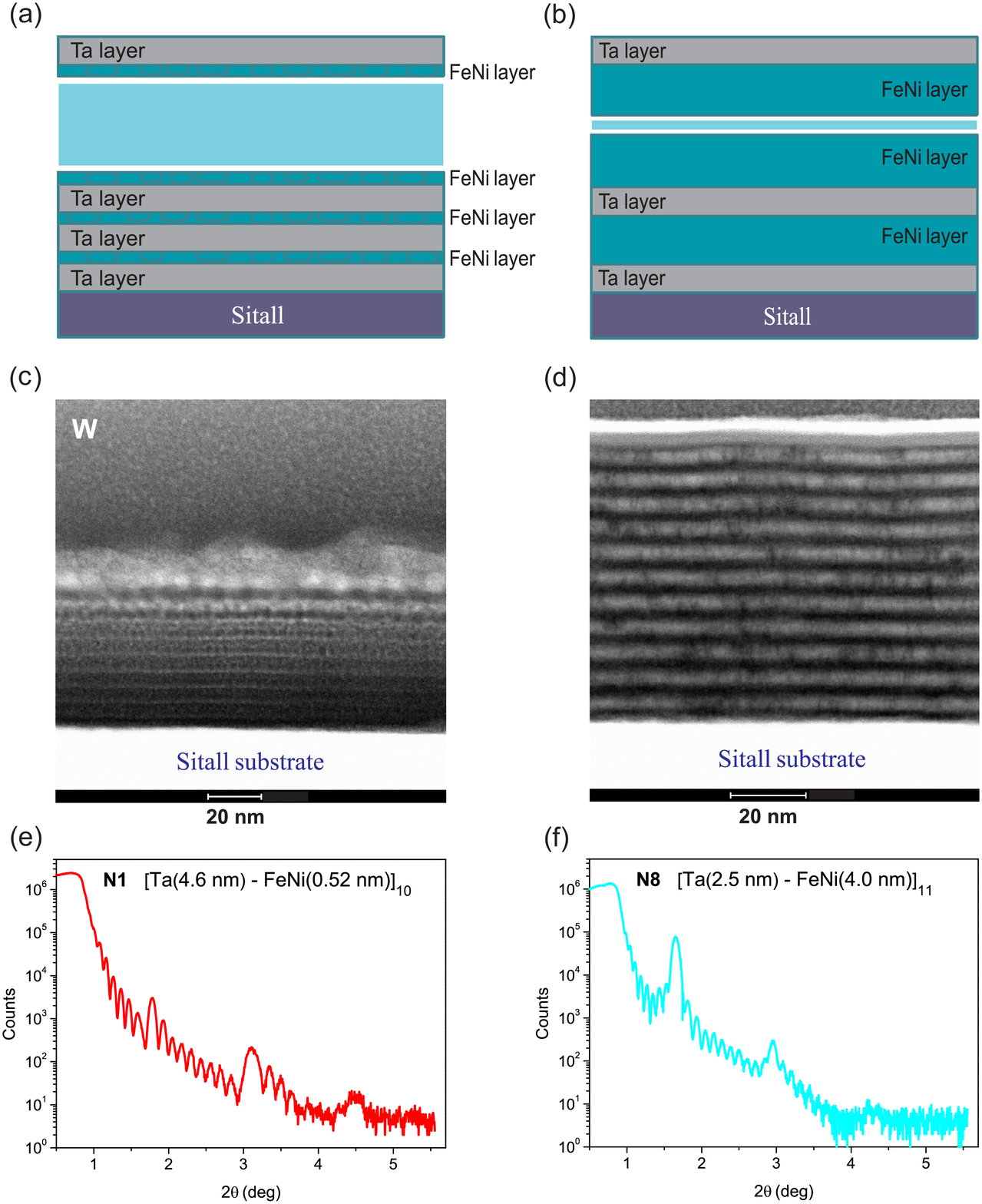}
\caption{{\bf STEM and x-ray reflectivity characterisation of the (Ta--FeNi)$_{\rm N}$ MLF samples.} (a) For the first series of the MLFs, nominal thickness of the FeNi nanoisland layer was of 0.52\,nm and the thickness of the Ta layer varied in the studied samples N1(4.6 nm), N2(2.3 nm), and N3(1.2 nm), where the number of periods was N=10, 11, and 14, respectively. (b) For the second series of the MLFs, the Ta layer thickness was 2.5 nm and the thickness of the FeNi layer varied in the studied samples N5(1.0 nm), N6(1.6 nm), N7(2.0 nm), and N8 (4.0 nm), where the number of periods was N=11. (c,d) Bright-field STEM image from the Ta (dark gray colour) and FeNi (light gray colour) layers, and (e,f) x-ray reflectivity of the MLF samples N1 and N8, respectively.}
\label{SLscheme}
\end{figure}

The MLFs (Ta--FeNi)$_{\rm N}$--Ta were grown by alternating rf sputtering from 99.95\% pure Ta and Fe$_{21}$Ni$_{79}$ targets onto insulating Sitall-glass substrates (for more details, see Section Methods). 
Two series of the MLF samples were prepared. In the first series of the grown (Ta--FeNi)$_{\rm N}$--Ta/Sitall film samples (hereinafter referred to as N1, N2, and N3), the nominal thickness of the FeNi layer was 0.52 nm; the thickness of the Ta layer was of 4.6, 2.3, and 1.3 nm; and the number of layers was N = 10, 11, and 14, respectively (as schematically shown in Fig.\,\ref{SLscheme}(a)). Here the thickness of the FeNi layer was chosen to be 0.52 nm, when the layer is discontinuous and represented by single-domain FM nanoislands \cite{Sherstnev}. In the second series of the grown (Ta--FeNi)$_{\rm N}$--Ta/Sitall film samples (hereinafter referred to as N5, N6, N7, and N8, shown by schematics in Fig.\,\ref{SLscheme}(b)), the number of layers was N = 11, the Ta layer thickness was fixed at 2.5 nm, and the FeNi layer thickness was 1.0, 1.5, 2.0, and 4.0 nm, respectively, increasing above the FeNi film percolation threshold around 1.5--1.8\,nm \cite{Sherstnev,Pudonin_JETP}. Figure \ref{SLscheme}(c,d) shows 
bright-field scanning/transmission electron microscopy (STEM) images 
from the Ta and FeNi layers in the cross-section of the MLF samples N1 
and N8. The nominal Ta and FeNi thickness values are in good agreement 
with their values estimated from the cross-section STEM images. One 
can see that the STEM images reveal well defined interfaces in 
the MLF structures. The X-ray reflection from the representative 
MLF samples from the first and second series (N1 and N8) exhibits 
superlattice peaks (Kiessig's oscillations), as one can see from 
Fig.\,\ref{SLscheme}(e,f), respectively, indicating a periodic 
compositional modulation along the film growth direction. The observed 
Kiessig's oscillations resolved up to the third order give evidence 
of the relatively small interface roughness (see more details about 
the MLFs characterisation in the Supplementary Information for 
this paper).\\

\begin{figure}[b]
\includegraphics*[width=9.5cm]{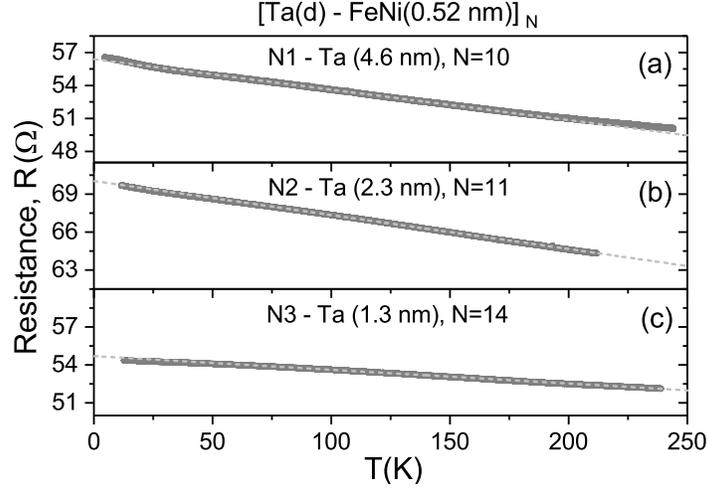}
\caption{{\bf I. Temperature-dependent dc transport for the Ta--FeNi MLFs.} (a-c) The dc resistance $R$ of the MLF samples N1--N3, respectively (see Fig.\,\ref{SLscheme}(a,c,e)), approximated by the linear dependence in the 80 -- 180 K temperature range, represented by the dashed lines.}
\label{DCN123}
\end{figure}

\begin{figure}[b]
\includegraphics*[width=9.5cm]{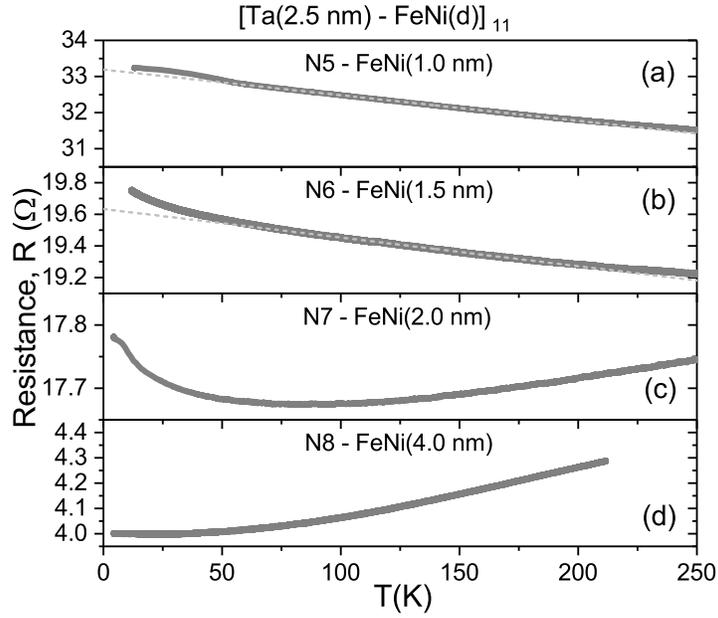}
\caption{{\bf II. Temperature-dependent dc transport for the Ta--FeNi MLFs.} (a-d) The dc resistance $R$ of the MLF samples N5--N8, respectively (see Fig.\,\ref{SLscheme}(b,d,f)), approximated by the linear dependence for the MLF samples N5 and N6 in the 80--180 K temperature range, as indicated by the dashed lines.}
\label{DCN5678}
\end{figure}

\hspace{-1.2em}{\large \bf Results}\\
\hspace{-1em}{\bf Temperature-dependent dc transport of the Ta--FeNi multilayer films}\\
Figures\,\ref{DCN123}(a-c) and \ref{DCN5678}(a-d) show the temperature dependence of the dc resistance of the MLF samples N1--N3 and N5--N8 measured on heating from the low temperature of 5 K to about 200--250 K. From Fig.\,\ref{DCN123}(a-c) one can see that the dc resistance of the MLF samples N1, N2, and N3 decreases by about 7--15\% with increasing temperature demonstrating non-metallic ($d R/dT<0$) character. We found that in the 80\,--\,180 K temperature range the MLFs resistance variation can be well approximated by the linear temperature dependence characterised by the negative slope values, as it is shown in Fig.\,\ref{DCN123}(a-c).

\begin{figure}[b]
\includegraphics*[width=10.5cm]{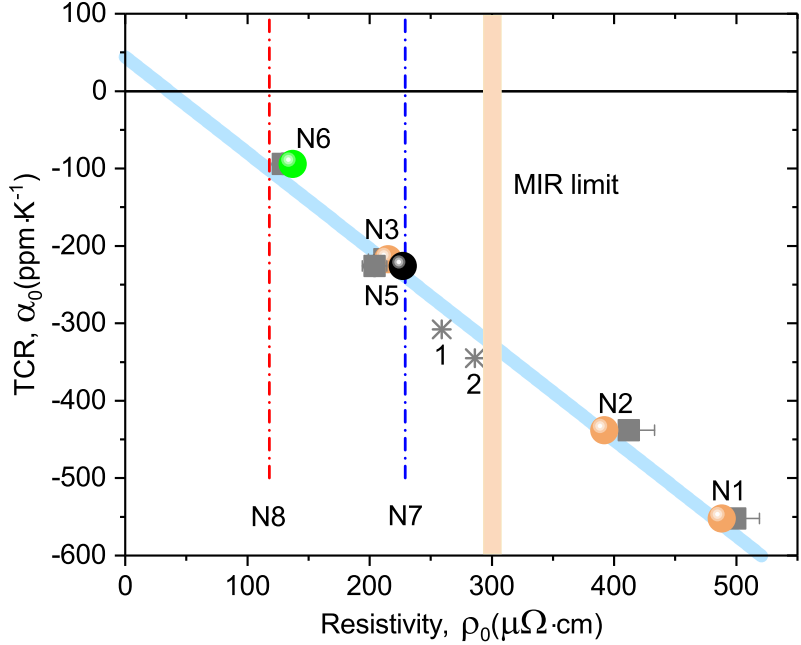}
\caption{{\bf Mooij correlations ($\alpha_0$ vs $\rho_0$) for the Ta layer in the Ta--FeNi MLFs from the dc transport and optical studies.} 
The results of the dc transport study ($\alpha_0$ vs $\rho_0$) for (i) the MLF samples N1--N3, N5, and N6 are represented by square symbols (see also Table\,\ref{table1}), and for (ii) the 50- and 33-nm thick single-layer Ta films are shown by asterisks marked as 1 and 2, respectively \cite{Kovaleva_metals}. The results of the optical study are represented by ($\alpha_0$ vs 1/$\sigma_{1(\omega\rightarrow0)}$) and include the optical Drude dc limit resistivities (1/$\sigma_{1(\omega\rightarrow0)}$) for the Ta layer in the MLFs N1--N3 (orange circles) \cite{Kovaleva_APL_2} and N5 (black circle) and N6 (green circle) (see also Fig.\,\ref{Sigma}(a,b) and Table\,\ref{table2}). The dash-dotted lines correspond to the optical Drude dc limit resistivities of the Ta layer in the MLFs N7 and N8, where the $\alpha_0$ values were not possible to extract from the present dc transport measurements (see Eq.\,(\ref{DCR}) and the text). Notice good agreement between the dc transport and optical dc limit resistivity values. }
\label{TCR}
\end{figure}

Following the representation used in the Mooij plot \cite{Mooij} for disordered metallic systems, $R=R_0\left[ 1+\alpha_0(T-T_0)\right]$, we determine the associated TCR ($\alpha_0$) and $R_0$ (interpolated to $T_0=298$\,K) values, which are listed in Table\,\ref{table1}. Suggesting that the discontinuous nanoisland FeNi layer is non-conducting in the MLFs N1, N2, and N3, the  resistivity $\rho_0$ values of the Ta layer were estimated using the equation $\rho_0=$N$R_0\gamma d$, where N is the number of layers, $d$ is the nominal Ta layer thickness, and $\gamma=c/a$ is a geometrical factor determined by the ratio of the sample width to the distance between the potential contacts. The resulting $\alpha_0$ and $\rho_0$ values are listed in Table\,\ref{table1} and displayed in the Mooij plot coordinates in Fig.\,\ref{TCR}.

We would like to note that the estimated dc conductivity values $\sigma_0$=1/$\rho_0$ (see Table\,\ref{table1}) are in good agreement with the optical dc limit $\sigma_{1\,(\omega\rightarrow0)}$ of the Ta layer Drude response in the MLF samples N1, N2, and N3 of $\simeq$ 2050 (4.6 nm), $\simeq$ 2550 (2.3 nm), and  $\simeq$ 4650 (1.3 nm) $\Omega^{-1}\cdot$cm$^{-1}$ (depicted in Fig.\,{\ref{TCR}}), resulting from our recent spectroscopic ellipsometry study \cite{Kovaleva_APL_2}.

The obtained data ($\alpha_0$, $\rho_0$) for the Ta layer of different thickness in the MLFs match the range of the Mooij plot for disordered metals and alloys \cite{Mooij,Tsuei}. In particular, the ($\alpha_0$, $\rho_0$) values for the Ta layer in the sample N1 fall in the area around ($-500$\,ppm/K, 500\,$\mu \Omega\cdot$cm) on the Mooij plot, seemingly indicating its unusual transport properties. Indeed, here the dc resistivity of the Ta layer is well above the critical value $\rho^*\sim$\,300\,$\mu\Omega\cdot$cm of the MIR limit (Eq.\,(\ref{MIR})) for disordered metals. The unforeseen result obtained is that the negative slope decreases in the investigated series of the MLF samples N1, N2, and N3 upon decreasing the Ta layer thickness (see Table\,\ref{table1} and Fig.\,\ref{TCR}). However, one would naturally expect that the absolute value of negative TCR will increase with degree of disorder in the thinner Ta layer. This is illustrated by the ($\alpha_0$,$\rho_0$) data determined for the 50- and 33-nm thick single-layer Ta films grown onto the Sitall substrate at the same rf sputtering conditions \cite{Kovaleva_metals} (see Fig.\,\ref{TCR}). Thus, we can conclude that the $\rho_0$ and TCR absolute values in the MLF samples N1, N2, and N3 preferentially depend on and are determined by the Ta layer thickness (and so the magnetic interaction strength). Following the Mooij rule \cite{Mooij}, the TCR value 
tends to be more negative with increasing $\rho_0$ 
(as demonstrated by Fig.\,\ref{TCR}).

Figure\,\ref{DCN5678}(a-d) clearly demonstrates the evolution of the dc transport in the MLF samples of the second series from the non-metallic ($d R/dT<0$ in N5 and N6) to metallic ($d R/dT>0$ in N7 and N8) character with an increase in the FeNi layer thickness from 1.0, 1.5, 2.0 to 4.0 nm. Naturally, this behaviour can be associated with the FeNi layer, which undergoes an insulator-to-metal transition due to appreciable coalescence of FeNi nanoislands in the layer with the thickness above the percolation threshold at the critical thickness $d_c$\,$\simeq$\,1.5--1.8\,nm \cite{Sherstnev,Pudonin_JETP}. We found that below the percolation threshold of the FeNi film the resistance  variation in the MLF samples N5 and N6 can be adequately approximated by the linear dependence in the 80\,--\,180\,K temperature range  (see Fig.\,\ref{DCN5678}(a,b)) characterised by the negative slope values. Given that the discontinuous nanoisland FeNi layer is non-conducting below the percolation threshold in the samples N5 and N6, we evaluated the ($\alpha_0$, $\rho_0$) values for the Ta layer in these MLF samples (see Table\,\ref{table1} and Fig.\,\ref{TCR}). Here, the Ta layer dc resistivity appeared to be below the MIR limit ($\rho^*\sim$\,300\,$\mu\Omega\cdot \rm{cm}$) and, following the Mooij rule, their TCR values tend to be less negative (see Fig.\,\ref{TCR}). 
We would like to note that the $R(T)$ dependencies shown in  Figs.\,\ref{DCN123}(a) and \ref{DCN5678}(a,b) indicate insignificant deviations from the linear approximations above 180\,K, associated with the dc transport peculiarities, which require a more comprehensive study at elevated temperatures.

At and above the percolation threshold of the FeNi film at the critical thickness $\sim$\,1.5--1.8\,nm \cite{Sherstnev,Pudonin_JETP}, the $R(T)$ behaviour of the MLF samples N7 and N8 is determined by the cumulative contributions from the inclusive FeNi and Ta layers. The $R(T)$ dependence of the MLF sample N7 exhibits a clearly pronounced minimum at $\sim$\,80 K (see Fig.\,\ref{DCN5678}(c)), whereas it is very weakly pronounced and shifted to the lower temperature of $\sim$\,20 K for the sample N8 (omitted on the scale of Fig.\,\ref{DCN5678}(d)). Above the minima, we observed a rise of the  $R(T)$ dependence for the samples N7 and N8 with metallic ($d R/dT>0$) character (see Fig.\,\ref{DCN5678}(c,d)). We found that here the temperature resistance  variation can be well approximated by the $T^2$ dependence (omitted), which can be naturally ascribed to inelastic electron-electron scattering \cite{Abrikosov} in the percolating and continuous FeNi layers. Here, the Ta and FeNi layers having the nominal thickness $d$ and $h$, respectively, are both conducting. The resistance $R$ of the MLF sample can be determined by the following relationship  
\begin{eqnarray}
\frac{1}{R\rm N\gamma} =\left( \frac{h}{\rho_0} \right)^{FeNi}+\left( \frac{d}{\rho_0}\right)^{Ta} =\left(h\sigma_0\right)^{FeNi}+\left(d\sigma_0 \right)^{Ta}
\label{DCR}
\end{eqnarray}
expressed via  the resistivities $\rho_0$ (or conductivities 
$\sigma_0$\,=\,$1/\rho_0$) of the inclusive layers. However, since in 
Eq.\,(\ref{DCR}) we have two unknown values $\rho_0^{Ta}$ 
and $\rho_0^{FeNi}$, it is not possible to attain the Ta layer dc conductivity properties in the MLF samples N7 and N8 via the dc transport measurements. Here, we determine the optical conductivity properties of the MLF 
samples N5--N8 from the spectroscopic ellipsometry study 
presented below.\\

\vspace{0.3cm}
\hspace{-1.2em}{\bf Spectroscopic ellipsometry of the Ta--FeNi multilayer films}\\
In the recent publication \cite{Kovaleva_APL_2}, we reported results of our wide-band spectroscopic ellispometry study (in the 0.8--8.5 eV spectral range) of the MLF samples (Ta--FeNi)$_{\rm N}$ from the first series (as schematically presented in Fig.\,\ref{SLscheme}(a)). The Ta layer complex dielectric function was obtained from the Drude-Lorentz simulations, which revealed different metallicity characters in the MLF samples N1--N3. This strongly 
suggests that the itinerant charge carrier response in the Ta 
intralayer preferentially depends on the distance between the 
adjacent 0.52-nm thick nanoisland FeNi layers and so determined 
by the magnetic interaction strength between them.

\begin{figure}[b]
\includegraphics*[width=12.0cm]{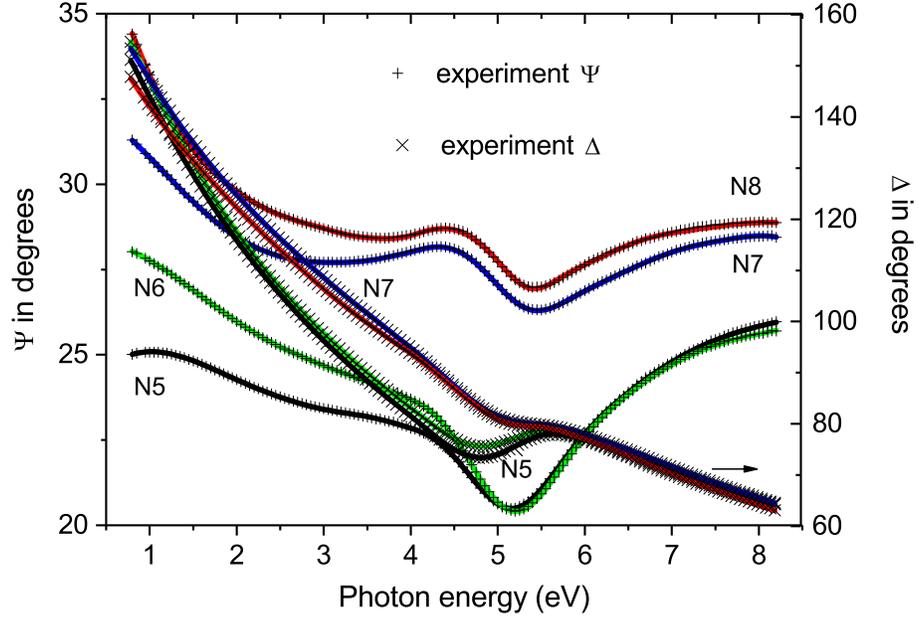}
\caption{{\bf Spectroscopic ellipsometry data.} Ellipsometric angles, $\Psi(\omega)$ and $\Delta(\omega)$, measured for the MLF samples N5--N8 at the angle of incidence 70$^\circ$ (displayed by the symbols). The solid lines show the fitting results using the Drude-Lorentz model (Eq.\,(\ref{DispAna})) in the simulation of the MLFs response.}
\label{PsiDelta}
\end{figure}

Here, using the  spectroscopic ellipsometry approach, we focused our research on the MLF samples  (Ta--FeNi)$_{\rm N}$ from the second series (as schematically shown in Fig.\,\ref{SLscheme}(b)) in the limit where the FeNi layer thickness is varied across the percolation threshold around 1.5--1.8\,nm \cite{Sherstnev,Pudonin_JETP}. The ellipsometric angles $\Psi(\omega)$ and $\Delta(\omega)$ were measured at room temperature at two angles of incidence of 65$^\circ$ and 70$^\circ$ (see Fig.\,\ref{PsiDelta}). The measured ellipsometric angles $\Psi(\omega)$ and $\Delta(\omega)$ were fitted in the framework of the multilayer model [Ta(2.5\,nm)--FeNi($h$)]$_{\rm 11}$--Ta(2.5\,nm)/Sitall (where $h$\,=\,1.0,\,1.5,\,2.0,\,and\,4.0\,nm) using the J.A. Woollam VASE software \cite{VASE software}. The complex dielectric function $\tilde \epsilon(\omega)=\varepsilon_1(\omega)+\rm{i} \varepsilon_2(\omega)$ of each layer was modeled by a Drude term, which is a zero-resonance energy Lorentz oscillator used to represent free charge carriers,  and a sum of contributions from higher-energy Lorentz oscillators
\begin{eqnarray}
\tilde\varepsilon(E\equiv \hbar\omega)=\epsilon_{\infty}-\frac{A_D}{E^2+{\rm i}E\gamma_D}+\sum_j\frac{A_j \gamma_jE_j}{E_j^2-E^2-{\rm i}E\gamma_j}, 
\label{DispAna}
\end{eqnarray}
where $\varepsilon_{\infty}$ is the core contribution to the dielectric function. The adjustable (fitting) Drude parameters were $A_D$ (which is related to the plasma frequency $\omega_p$ via $A_D=\varepsilon_{\infty}\hbar\omega^2_p$) and scattering rate $\gamma_D$. Each Lorentz oscillator was fitted with three adjustable parameters $E_j$, $\gamma_j$, and $A_j$ of the peak energy, the full width at half maximum, and the $\varepsilon_2$ peak height, respectively. 

The Ta and FeNi layers in each MLF structure were described by different dispersion models (Eq.\,(\ref{DispAna})), including the Drude term and Lorentz oscillators. In the simulation of the ellipsometry data, the discontinuous nanoisland FeNi layers were represented by the effective dielectric function in the effective medium approximation (EMA). To the utilized multilayer model, the complex dielectric function spectra of the blank Sitall substrate obtained from our complementary ellipsometry measurements, were substituted. The Ta and FeNi layer thicknesses were fitted to their respective nominal values. The high quality of the fit is demonstrated by Fig.\,\ref{PsiDelta}, where we present the measured ellipsometric angles $\Psi(\omega)$ and $\Delta(\omega)$ along with the fitting results. The quality of the fit was verified by the coincidence within the specified accuracy of less than 5\% with the ellipsometric angles $\Psi(\omega)$ and $\Delta(\omega)$ measured at two angles of incidence of 65$^\circ$ and 70$^\circ$. It was verified that the simulation in the framework of the multilayer model (Ta--FeNi)$_{\rm N}$--Ta/Sitall, where the Ta--FeNi interface roughness is explicitly included, does not improve the fit (for more details, see our recently published study \cite{Kovaleva_APL_2}). In particular, we expect that the Ta--FeNi interface roughness is essentially incorporated in the effective dielectric function of the nanoisland FeNi layer.

From the multilayer model simulations by using the dispersion model introduced by Eq.(\ref{DispAna}), the imaginary and real parts of the dielectric function spectra, $\varepsilon_2(\omega)$ and $\varepsilon_1(\omega)$, of the FeNi and Ta layers in the MLF structures N5--N8 were extracted, which are displayed in Figs.\,\ref{EPSFeNiTa}(a-d) (for more details of the dispersion analysis, see the Supplementary Information). The EMA dielectric function of 
the FeNi layers in the MLF samples N5 and N6 was modeled by three 
Lorentz oscillators, while any Drude contribution was vanished 
out during the fit, thus notifying that the optical dc limit 
$\sigma^{FeNi}_{1\,(\omega\rightarrow0)}\approx0$. This gives more confidence to our estimates of the Ta layer resistivity $\rho_0$ in the MLF samples N5 and N6 (see Table\,\ref{table1}) from the present dc transport measurements relied on the guess that the discontinuous FeNi layers are non-conducting in these structures. From Fig.\,
\ref{EPSFeNiTa}(a) one can clearly see that the $\varepsilon_2(\omega)$ function of the FeNi layer dramatically increases at low probed photon energies with increasing the FeNi layer thickness above the percolation threshold. Simultaneously, the $\varepsilon_1(\omega)$ function of the FeNi layer progressively decreases with increasing the FeNi layer thickness and for the 4nm-thick FeNi layer it exhibits a sharp downturn to negative values at the lowest probed photon energies (see Fig.\,\ref{EPSFeNiTa}(b)), demonstrating the metallic behaviour. The dielectric function of the FeNi layers in the MLF samples N7 and N8 was modeled by Drude resonance and four higher-energy Lorentz oscillators. According to our simulation results, we expect that the Drude resonance becomes rather narrow in the sample N8 well above the percolation transition, whereas it is quite wide  at the percolation threshold in the sample N7. The resulting values of the Drude dc conductivity limit $\sigma^{FeNi}_{1\,(\omega\rightarrow0)}$  and scattering rate $\gamma^{FeNi}_D$  in the MLF structures N5--N8 are given in Table\,\ref{table2}.

From Fig.\,\ref{EPSFeNiTa}(c,d) one can follow the trends of the foreseen behaviour in the complex dielectric function spectra of the 2.5-nm thick Ta intralayer accompanying the insulator-to-metal transition within the FeNi layer, resulting from the current Drude-Lorentz model simulations of the MLF samples N5--N8. Note that the metallicity character of the Ta layer changes in the MLF sample series N5--N8 across the FeNi film percolation transition, although not sequentially. Indeed, the displayed $\varepsilon_2(\omega)$ and $\varepsilon_1(\omega)$ spectra indicate that, in agreement with the dc transport measurements (see Table\,\ref{table1} and Fig.\,{\ref{TCR}}) the Ta layer in the sample N6 reveals better  metallicity properties than that in the sample N5. The dielectric function response of the Ta layer in the sample N7 suggests its poorer metallicity properties than in the sample N6, whereas the best metallicity properties in the whole MLF sample series N5--N8 are exhibited by the Ta layer in the sample N8.

\begin{figure}[b]
\includegraphics*[width=7.5cm]{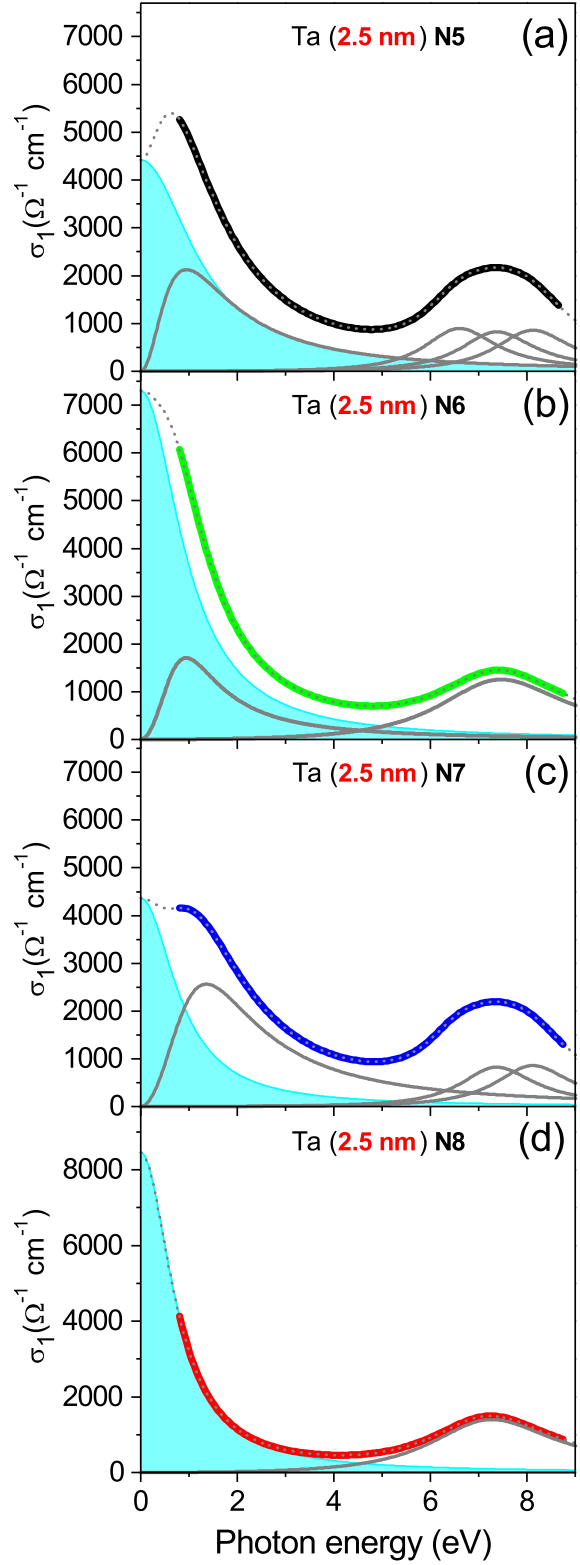}
\caption{{\bf Wide-range optical conductivity of the Ta layer in the MLFs Ta--FeNi.} (a-d) The Ta layer optical conductivity $\sigma_1(\omega)=\frac{1}{4\pi}\omega\varepsilon_2(\omega)$ in  the MLF samples N5, N6, N7, and N8 shown by solid black, green, blue, and red curves, respectively.  The contributions from the Ta layer Drude term is shown by the cyan shaded area and from the Lorentz oscillators by solid gray lines. The agreement of the dispersion analysis with the $\sigma_1(\omega)$ dependence is demonstrated by the summary Drude-Lorentz contribution, shown by dotted lines.}
\label{Sigma}
\end{figure}

Figure\,\ref{Sigma}(a--d) displays the evolution of the Ta intralayer optical conductivity $\sigma_1(\omega)$\,=\,$\frac{1}{4\pi}\omega\varepsilon_2(\omega)$ upon increasing the FeNi layer thickness $h$ across the percolation threshold in the MLF structures [Ta(2.5\,nm)--FeNi($h$)]$_{11}$ for the samples N5--N8, respectively. Here, the contributions from the Drude term and Lorentz oscillators resulting from the multilayer model simulations using Eq.\,(\ref{DispAna})  are explicitly demonstrated. The corresponding values of the optical Drude conductivity limit $\sigma^{Ta}_{1\,(\omega\rightarrow0)}$  and scattering rate $\gamma^{Ta}_D$ are presented in Table\,\ref{table2}.
From Fig.\,\ref{Sigma}(a--d) one can see that the Ta layer optical conductivity $\sigma_1(\omega)$ in the MLF structures N5--N8 can be well described in the framework of the elaborated Drude-Lorentz model.\\

\vspace{0.3cm}
\hspace{-1.2em}{\large \bf Discussion}\\               
Let us follow the trends occurring in the Ta layer optical conductivity for the samples N2 \cite{Kovaleva_APL_2}, the 33-nm thick single Ta layer \cite{Kovaleva_APL_1,Kovaleva_metals}, N5, and N6, which span a major part of the ($\alpha_0$,\,$\rho_0$) Mooij plot related to the negative TCR values (see Fig.\,\ref{TCR}). According to the results obtained from the dc transport study (see Table\,\ref{table1}), the Ta layer in the sample N6 (Ta\,=\,2.5\,nm, FeNi\,=\,1.5\,nm) exhibits the best dc conductivity properties with $\alpha_0\sim-$\,95\,ppm/K and $\rho_0\sim$124\,$\mu\Omega$$\cdot$cm$^{-1}$. For the chosen sample series, the Ta layer in the sample N2 (Ta\,=\,2.3\,nm, FeNi\,=\,0.52\,nm) demonstrates the worst conductivity properties with $\alpha_0\sim-$\,438\,ppm/K and $\rho_0\sim$\,412\,$\mu\Omega$$\cdot$cm$^{-1}$. The 33-nm single-layer Ta film displays the dc conductivity properties peculiar to  the MIR limit. Indeed, according to the dispersion analysis using Eq.\,(\ref{DispAna}), the Ta layer low-energy response is dominated by the Drude term demonstrating the optical dc conductivity limit $\sigma^{Ta}_{1\,(\omega\rightarrow0)}\sim$\,3300\,$\Omega^{-1}\cdot$cm$^{-1}$ (in good agreement with Eq.\,(\ref{MIR})) and the intense low-energy Lorentz band peaking at 2.2\,eV \cite{Kovaleva_APL_1}. What are the main trends observed in the optical conductivity spectra of these sample on moving along the ($\alpha_0$,$\rho_0$) line (see Fig.\,\ref{TCR}) in the opposite directions from the MIR limit? Analyzing the optical conductivity spectra shown in Fig.\,\ref{Sigma}(a-d) and in our previous papers (see Fig.\,4(d)\cite{Kovaleva_APL_1,Kovaleva_APL_2}) one can see the following systematic changes (i) the Drude dc conductivity decreases below the MIR limit and the low-energy band shifts to a higher energy of 2.7\,eV in the sample N2, (ii) the Drude dc conductivity increases above the MIR limit and the low-energy band shifts to the lower energy about 1 eV in the samples N5 and N6, and (iii) the intensity of the low-energy band decreases on  the conductivity enhancement in the sample N6. According to our simulations, the best metallicity properties among the whole MLF sample series N5--N8 are demonstrated by the Ta layer in the sample N8 (see Table\,\ref{table2} and Fig.\,\ref{Sigma}(d)), where the low-energy band completely 
disappears from the optical conductivity spectrum.

Near the percolation transition within the FeNi layers strong magnetic dipole-dipole and exchange interactions between the FM nanoislands become relevant. It was shown that in quasi-2D systems of FM nanoparticles collective superferromagnetic (SFM) states can exist in their self-assembled local arrangements (clusters) at comparatively high temperatures \cite{Kleemann,Kovaleva_JNM,Kovaleva_JNR}. We suppose that in here GMR-like case is realized, where the interlayer coupling between two neighboring FeNi layers is driven by the indirect exchange interactions via the itinerant charge carriers of the Ta spacer layer. 
Their kinetic energy enhancement in the regime of strong FM 
correlations between giant magnetic moments of FeNi nanoislands 
(similar to the double-exchange (DE) model used in the description 
of the colossal magnetoresistance (CMR) effect) may lead to the delocalisation of electrons from the associated localised electronic states.
As a result, the Drude contribution increases implying the delocalisation of charge carriers simultaneously with the intensity decrease of the  low-energy excitation band. According to the obtained results, this may lead to the substantial Ta layer resistivity decrease of about 60\% (normalized to the MIR limit value  $\rho^*\sim$\,300\,$\mu\Omega\cdot$cm), introduced by 
the magnetic interaction between the FeNi layers.

Another  paradigm is represented by the MLF sample N1, where the 0.52-nm thick FeNi layer is far from the percolation transition and the FeNi layers are well separated from each other (by 4.6 nm). Earlier, we established that ultrathin monolayers of FeNi nanoislands exhibit superparamagnetic-like (SPM) behaviour at room temperature, associated with their giant magnetic moments of  10$^3$--10$^5$ $\mu_{\rm B}$ \cite{Kovaleva_JNM,Kovaleva_JNR}. In the MLF sample N1, incorporating such FeNi nanoisland layers, the Drude dc conductivity drops well below the MIR limit (see Table\,\ref{table1}) and the low-energy band at 2 eV becomes essentially suppressed, while the pronounced increase in the higher-energy band around 6--8 eV is observed (see Fig.\,4(e) from our previous article \cite{Kovaleva_APL_2}), which can be associated with new localised electronic states. According to the obtained results, these new localisation phenomena introduced by an additional strong magnetic disorder and long-range many-body interactions between giant magnetic moments of FeNi nanoislands via the remaining free itinerant charge carriers of  a disordered metal by  means of RKKY-type  indirect exchange, lead to the additional Ta layer resistivity increase (normalized to the MIR limit value  $\rho^*\sim$\,300\,$\mu\Omega\cdot$cm) up to 60\%.

To shed light on our experimental findings, 
below we present a simple model of giant magnetic moments 
(superspins $\textbf{S}_i=(S_{xi},S_{yi},S_{zi})$) 
of FM FeNi nanoislands interacting via indirect exchange mediated 
by conduction electrons of Ta layers. This model can be described by the following Hamiltonian 
\begin{equation}
 \label{eq:1} H_0 =
\sum\limits_{ i,j} J( k_{\rm F} r_{ij})(\textbf{S}_i \cdot \textbf{S}_j),
\end{equation}
where $k_{\rm F}$ is the Fermi momentum of the Fermi sea associated with the conduction electrons of the Ta nanolayer, and 
$J( k_{\rm F} r_{ij})$ is responsible for the interaction between 
the superspins $\textbf{S}_i$ and $\textbf{S}_j$ separated by 
the distance $r_{ij}$, which occurs due to Friedel-like 
oscillations of electron spin density induced by the spin 
polarization of an electron cloud nearby a FM FeNi nanoisland.
Such polarization arises due to exchange forces between localised magnetic moments of the FM FeNi nanoislands and mobile conduction electrons of 
the Ta nanolayer. The suggested interaction model has a certain  
similarity to the RKKY interaction \cite{Ruderman,Kasuya,Yoshida}. 
The system under study has a quasi-two-dimensional character 
implied by the structure of Ta/FeNi nanolayers. Due to this 
quasi-two-dimensionality, the long-range interaction between 
superspins $\textbf{S}_i$ and $\textbf{S}_j$ of the FeNi nanoislands 
separated by the distance $x$ takes the form~\cite{Klein1975}
  \begin{equation}
   \label{eq:2}
J(x) \sim \frac{n^2 J^2}{E_F}\frac{\sin{x}}{x^2},
 \end{equation}
where $n$ is the electron density per Ta atom, $J$ is the exchange 
interaction between electron spins of the Ta subsystem and superspins of FeNi nanoislands.

In our approach, we used second-order perturbation theory to describe 
an indirect exchange coupling whereby the spin of one Fe or Ni atom 
of a FeNi nanoisland interacts with a conduction electron of Ta, which 
then interacts with another spin of different FeNi nanoisland. 
Making summation over all atoms of one FeNi nanoisland, we will 
have interaction of the Ta conduction electron with the superspin of this FeNi nanoisland. Doing the same summation for the second FeNi nanoisland, we can find out a correlation energy between the two superspins associated with two different magnetic FeNi nanoislands. Thereby, the superspins 
produce a pronounced effect on the Fermi sea of electrons in the Ta 
nanolayer. When the Ta layer thickness is fixed (as for the Ta--FeNi MLFs from the second series), and so the distance between neighboring 
FeNi layers is fixed at 2.5\,nm, the interaction in Eq.\,\ref{eq:1} will be ferromagnetic \cite{Huetten} and will have a certain similarity to the double-exchange (DE) model, which is used, in particular, 
in description of the transport mechanism and optical conductivity 
properties of colossal magnetoresistive (CMR) manganites \cite{Millis,Boris}.  

When the FeNi layer is discontinuous and represented by randomly distributed distant FM FeNi nanoislands, this interaction will oscillate from ferromagnetic to the antiferromagnetic one with the period, which is very small, about a few nanometers. Therefore, the strong interaction between superspins oscillates between ferromagnetic and antiferromagnetic types as a function of the distance between the FeNi nanoislands. Note that even the size of these islands can be comparable with the period of such oscillations. This creates a strong frustration in the system impeding finding its ground state. As a result, the system is slowly migrating from one metastable state to another reminding the situation characteristic of spin and orbital glass systems~\cite{KusmartsevPLA1992}. Such frustrations and strong interactions between superspins can lead to some kind of a non-ergodic state.

To summarize, using dc transport and wide-band (0.8--8.5 eV) spectroscopic ellipsometry experimental techniques, we investigated the conductivity properties of the Ta intralayer inside the MLF structures (Ta--FeNi)$_{\rm N}$ in two limits of the magnetic layer thickness (i) when the FeNi layer has a discontinuous nanoisland structure and  (ii) when the FeNi layer thickness is increasing  across the percolation threshold. We found that for the nanoisland structure of the FeNi layer the Ta layer dc resistance temperature variation can be adequately approximated by the linear dependence in the 80\,--\,180\,K temperature range characterised by the negative TCR ($\alpha_0$) values peculiar of strongly disordered metallic systems \cite{Lee}. Following the Mooij rule \cite{Mooij}, the determined values of $\alpha_0$ and dc resistivity $\rho_0$ fit the linear dependence, where the TCR values tend to be more negative with increasing $\rho_0$ (see Fig.\,\ref{TCR}). 
We discovered that the dc transport of the Ta layer strongly depends on the structural properties of the FeNi layers around the metal-insulator 
percolation transition, which, in turn, determine their magnetic 
properties \cite{Kovaleva_JNR} and, thereby, the stregth of magnetic 
coupling between neighboring FeNi layers mediated by the RKKY-type interaction. The entanglement of the RKKY-type interaction with the itinerant 
electrons of the Ta spacer lies in the root of the observed Mooij correlations driven by the interplay between the strength of the magnetic interaction and localisation of electons in the disordered Ta--nanoisland FeNi MLF structures. According to our results, this may lead to a substantial Ta layer 
resistivity decrease of about 60\% (normalized to the MIR limit value 
$\rho^*\sim$\,300\,$\mu\Omega\cdot$cm). From the multilayer model simulations, we extracted the Ta and FeNi layer dielectric function response in the (Ta--FeNi)$_{\rm N}$ MLFs. Near the MIR limit the Ta layer dielectric function is represented by the Drude resonance due to the Ta layer free charge carriers superimposed with the low-energy excitations at around 2--4 eV. The low energy excitaions appear in evidence of strong electron correlations in the 
systems with strong electron correlations \cite{Kovaleva_PRL,Kovaleva_PRB_LMO,Kovaleva_PRB_YTO,Oles}. We suggest that electronic correlations accompany the localisation of electrons in clusters of electronic inhomogeneities.
We found that when the dc conductivity and the optical Drude dc conductivity limit consistently increase above the MIR regime the low-energy excitations exhibit the red shift to about 1 eV, their intensity decreases, and finally their fingerprints dissappear when the FeNi layer has the thickness well above the percolation threshold. The observed behaviour signals 
progressive delocalisation of electrons with increasing the indirect 
exchange coupling strength.

Thus, we discovered that the Ta layer resistivity critically depends 
on the properties of the FeNi layer, when its morphology changes 
from the discontinuous nanoisland to continuous structure across the percolation threshold, where the TCR and $\rho_0$ values obey the Mooij correlations. In the optical conductivity spectra, while approaching the MIR limit, 
the Drude contribution of charge carriers decreases, signaling the enhanced localization effects, which are accompanied by the appearance of the 
low-energy optical bands in the spectral range (1-2 eV) characteristic of electronic correlations.

On the contrary, we discovered that when the FeNi layer is represented by distant randomly distributed giant magnetic moments of magnetic FeNi nanoislands the additional localisation effects where the Ta layer normalized dc conductivity falls down below the MIR limit by about 60\% take place. The discovered phenomenon, which can be associated with a large-scale fluctuating potential of magnetic origin and lead to non-ergodicity and purely quantum localisation effects \cite{Basko,Nandkishore,Huse,Khemani,Lukin}, need to be further challenged theoretically and experimentally. The obtained results demonstrate the advances of the used spectroscopic ellipsometry approach, which allowed us to extract the dielectric function response of each  inclusive layer within the ultrathin MLF structures, and must be essential for understanding the physics of the Anderson localisation phenomena in magnetic multilayers and from the application view of the GMR effect.

\vspace{0.3cm}
\hspace{-1.2em}{\large \bf Methods}\\
The MLFs (Ta--FeNi)$_{\rm N}$--Ta were grown by alternating rf sputtering from 99.95\% pure Ta and Fe$_{21}$Ni$_{79}$ targets onto insulating Sitall-glass substrates. The actual substrate temperature during deposition was 80\,$^\circ$C. The base pressure in the chamber was 2$\times$10$^{-6}$ Torr, and Ar gas flow with a pressure of 6$\times$10$^{-4}$ Torr was used for the sputtering process. The Ta and FeNi layer nominal thickness was determined by the deposition time, and the deposition rate was about 0.067 nm/s. The sputtered MLFs were characterised by x-ray diffraction (XRD) analysis. The periodicity of the grown MLFs (Ta\,--\,FeNi)$_{\rm N}$ was characterised by low-angle x-ray reflectometry. The x-ray measurements were carried out on a Bede 200 Goniometer operating at 55 kV and 300 mA with Cu K$_\alpha$ radiation $\lambda$\,=\,0.154 nm produced by an x-ray Generator with a rotating  Rigaku RU 300 anode (for more details, see the Supplementary Information to this paper).

The microstructure of the MLFs (Ta--FeNi)$_{\rm N}$ was examined on a Titan 80-300 scanning/transmission electron microscope (STEM)  (FEI, US) operating at the accelerating voltage 300 kV and equipped with the spherical aberration corrector of an electron probe in the bright- and dark-field regimes. In the dark-field imaging a high-angle annular dark-field (HAADF) detector (Fischione, US) registering large-angle scattered electrons was used.  The cross-section specimens were prepared by the standard focus ion beam procedure on Versa 3D and Helios NanoLab 600i (Thermo Fisher Scientific, US) dual beam microscopes. The instruments were equipped with an Omniprobe micromanipulator  (Omniprobe, US) and gas injection systems for Pt and W deposition. At the first stage, 2-$\mu$m thick W capping layer was formed at the sample surface to protect from possible damages. The cuts were performed using a 30 keV Ga$^+$ ion beam.  To remove the amorphous layer at the final stage of the cutting, the Ga$^+$ ion beam energy was reduced to 2 keV. Energy dispersive x-ray (EDX) microanalysis was performed using a spectrometer (Phoenix System, EDAX, US). (for more details, see the Supplementary Information to this paper).

\vspace{0.3cm}
\hspace{-1em}{\bf Experimental approach}\\
The dc resistance of the rf-sputtered MLFs was measured in a standard four-probe configuration in the linear I$-$V  regime using lock-in amplifiers (MFLI, Zurich instruments) and a voltage controlled current source (CS580, Stanford Research Systems). Ohmic contacts to the MLF samples were made by soldering. The samples were mounted on a cold finger inside the home-made insert and evacuated to a low-pressure environment at room temperature. The dc resistance was measured on heating  in a wide temperature range of 5--250 K. To avoid noticeable temperature delay during the measurements, the temperature variation rate was adjusted at 3.5 K/min.

In our complementary optical study we used spectroscopic ellipsometry 
approach. The grown MLF samples (Ta--FeNi)$_{\rm N}$--Ta/Sitall 
were measured by spectroscopic ellipsometry in a wide photon energy 
range of 0.8--8.5 eV with a J.A. Woollam VUV-VASE Gen II spectroscopic 
ellipsometer. The ellipsometry probes were obtained on the prepared 
MLF samples at two angles of incidence of 65$^\circ$ and 
70$^\circ$ at room temperature, as well as on the blank Sitall 
substrate.

\hspace{-1em}

\vspace{0.5cm}
\hspace{-1.2em}{\large \bf Acknowledgments}\\
This work was partially supported by the Czech Science Foundation (Project No. 20-21864S), European Structural and Investment Funds and the Czech Ministry of Education, Youth, and Sports (Project No. SOLID21, CZ.02.1.01/0.0/0.0/16$_{-}$019/0000760), and by the Russian Foundation for Basic Research (projects Nos. 19-02-00509 and 20-02-00015). The work of N.N.K., A.B.M., A.B.D., L.N.O., and B.A.A. is carried out within the state assignment of the Ministry of Science and Higher Education of the Russian Federation (theme ``Physics of condensed matter: new materials, molecular and solid state structures for nanophotonics, nanoelectronics, and spintronics'').

\vspace{0.5cm}
\hspace{-1.2em}{\large \bf Author contribution}\\
N.N.K. and D.C. carried out the spectroscopic ellipsometry 
measurements and analysed the data, A.B.M., A.B.D., and L.N.O. 
carried out the dc transport measurements and analysed the data, 
I.N.T. obtained STEM images, O. Pacherova fulfilled x-ray analysis, 
I.A.S. and F.A.P. prepared the multilayer films. A.K. participated 
in the discussions. N.N.K., K.I.K. and F.V.K. wrote the manuscript. 
A.D., Y.L. and B.A.A. supervised the project.

\vspace{0.5cm}
\hspace{-1.2em}{\large \bf Additional Information}\\
\hspace{-1em}{\bf Supplementary Information} is linked to the online version
of the paper at http://www.nature.com/srep.

\hspace{-1em}{\bf Competing financial interests:} The authors declare no
competing financial interests.

\hspace{-1em}{\bf Correspondence} and requests for materials should be addressed to N.N.K.

\begin{table}
\caption{TCR ($\alpha_0$) and dc resistivity ($\rho_0$) values for the Ta layer of different thickness ($d$) in the MLFs [Ta($d$)\,--\,FeNi($h$)]$_{\rm N}$ below the FeNi layer percolation threshold. $R_0$ is the MLF sample resistance obtained from the linear interpolation to $T_0=$\,298 K (see the text).}
\begin{tabular}{lccccccc}
\hline\noalign{\smallskip}
N&$d$ & $h$&$\gamma$&$R_0$&$\alpha_0$ & $\rho_0$&$\sigma_0$ \\
&(nm) & (nm)&&($\Omega$)&(ppm/K) & ($\mu\Omega\cdot$cm)&($\Omega^{-1}\cdot$cm$^{-1}$) \\
\noalign{\smallskip}\hline\noalign{\smallskip}
N1 & 4.6 &0.52&2.25&48.2 & $-552\pm13$ & 499$\pm$20 & 2000$\pm80$ \\
N2 &2.3&0.52&2.63 &61.9 & $-438\pm5$ & 412$\pm21$ & 2430$\pm$120 \\
N3 & 1.3&0.52& 2.27&51.4 & $-217\pm2$ & 212$\pm13$ & 4720$\pm280$ \\
N5 &2.5&1.0&2.38 &31.1 & $-226\pm2$ & 205$\pm10$ & 4880$\pm$250 \\
N6 & 2.5&1.5& 2.45&19.1 & $-94\pm1$ & 129$\pm6$ & 7750$\pm390$ \\
\noalign{\smallskip}\hline
\end{tabular}
\label{table1}
\end{table}

\begin{table}
\caption{Optical dc limit $\sigma_{1\,(\omega\rightarrow0)}$ and scattering rate $\gamma_D$ of the Drude resonance in the FeNi and Ta layers for  the MLF samples N5--N8 [Ta($d$=2.5\,nm)\,--\,FeNi($h$)]$_{11}$, obtained from the present spectroscopic ellipsometry study.}
\begin{tabular}{lccccc}
\hline\noalign{\smallskip}
N & $h$&$\sigma^{FeNi}_{1\,(\omega\rightarrow0)}$&$\gamma^{FeNi}_D$ &$\sigma^{Ta}_{1\,(\omega\rightarrow0)}$ &$\gamma^{Ta}_D$\\
& (nm)&($\Omega^{-1}\cdot$cm$^{-1}$)&(eV) &($\Omega^{-1}\cdot$cm$^{-1}$)&(eV)\\
\noalign{\smallskip}\hline\noalign{\smallskip}
N5   & 1.0& 0 &--& 4400 &1.29\\
N6   & 1.5& 0 & --&7300 &1.00\\
N7   & 2.0& 5290 & 3.64& 4360& 0.86\\
N8   & 4.0& 9390& 0.39&8450&0.78 \\
\noalign{\smallskip}\hline\noalign{\smallskip}
\end{tabular}
\label{table2}
\end{table}

\begin{figure}[b]
\includegraphics*[width=16.0cm]{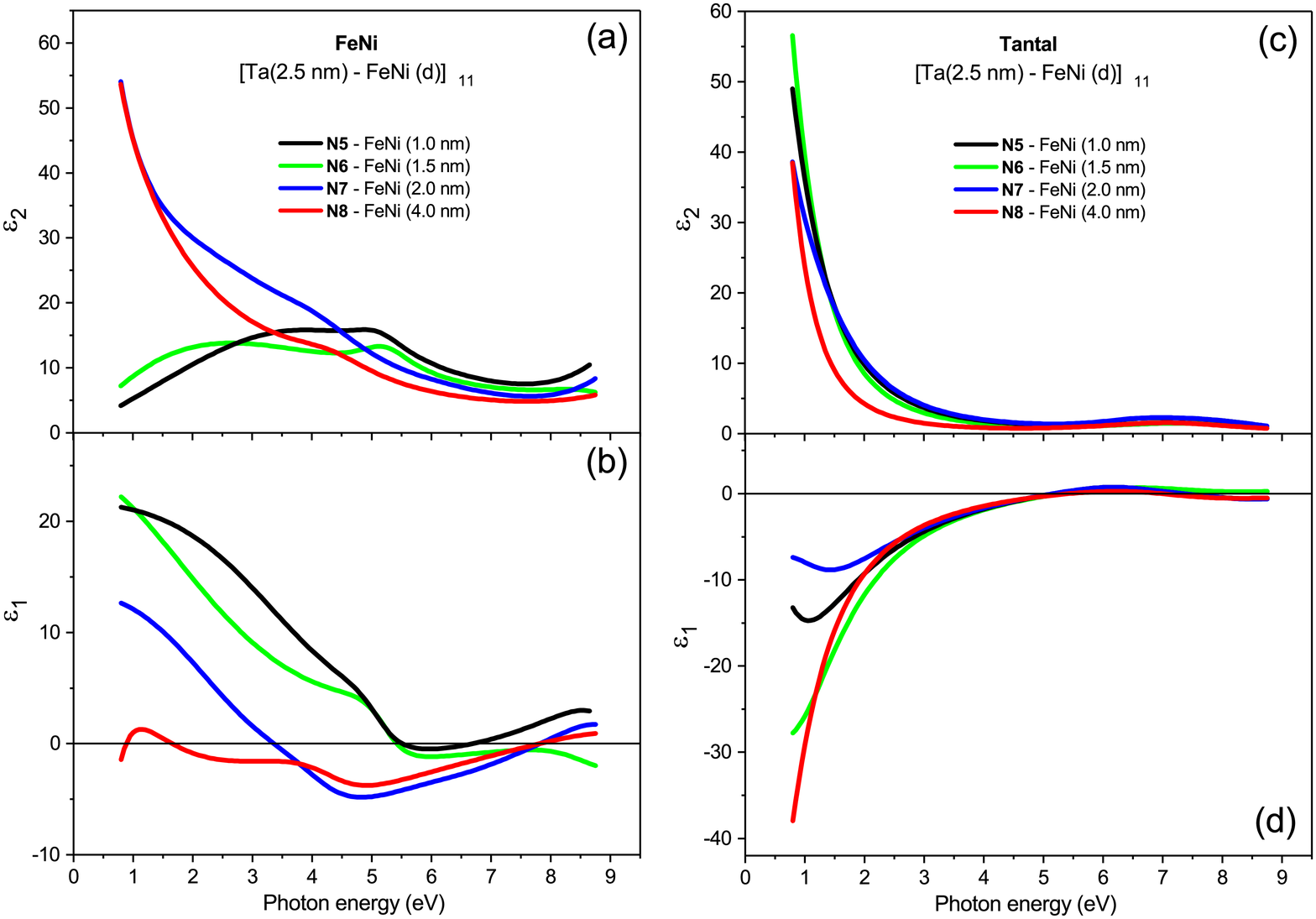}
\caption{{\bf Wide-range complex dielectric function response of FeNi and Ta layers in the studied Ta--FeNi MLFs.} (a,c) The $\varepsilon_2(\omega)$ and (b,d) $\varepsilon_1(\omega)$ dielectric function spectra of the FeNi and Ta layers, respectively, where the FeNi layer has different thickness of 1.0, 1.5, 2.0, and 4.0 nm in the MLF samples N5, N6, N7, and N8,  and the Ta layer thickness is regular and equal to 2.5 nm (shown by solid black, green, blue, and red curves, respectively).}
\label{EPSFeNiTa}
\end{figure}

\end{document}